# Engineering Annotations: A Generic Framework For Gluing Design Artefacts in Models of Interactive Systems

Marco Winckler, Université Côte d'Azur *

Philippe Palanque, Universite Paul Sabatier - Toulouse III, France

Jean-Luc Hack, Softeam, France

Eric Barboni, Universite Paul Sabatier - Toulouse III, France

Olivier Nicolas, Softeam, France

Laurent Gonçalves, Softeam, France

Along the design process of interactive system many intermediate artefacts (such as user interface prototypes, task models describing user work and activities, dialog models specifying system behavior, interaction models describing user interactions …) are created, tested, revised and improved until the development team produces a validated version of the full-fledged system. Indeed, to build interactive systems there is a need to use multiple artefacts/models (as they provide a complementary view). However, relevant information for describing the design solution and/or supporting design decisions (such as rational about the design, decisions made, recommendations, etc.) is not explicitly capturable in the models/artefacts, hence the need for annotations. Multi-artefacts approaches usually argue that a given information should only be present in one artefact to avoid duplication and increase maintainability of the artefacts. Nonetheless, annotations created on one artefact are usually relevant to other artefacts/models.  So that, there is a need for tools and techniques to coordinate annotations across artefacts/models which is the contribution of the present work. In this paper, we propose a model-based approach that was conceived to handle annotations in a systematic way along the development process of interactive systems. As part of the solution, we propose an annotation model built upon the W3C's Web Annotation Data Model. The feasibility of the approach is demonstrated by means of a tool suite featuring a plugin, which has been deployed and tested over the multi-artefacts. The overall approach is illustrated on the design of an interactive cockpit application performing two design iterations. The contribution brings two main benefits for interactive systems engineering: i) it presents a generic pattern for integrating information in multiple usually heterogenous artefacts throughout the design process of interactive systems; and ii) it highlights the need for tools helping to rationalize and to document the various artefacts and the related decisions made during interactive systems design.

## 1 INTRODUCTION

Design is a problem-solving process whose main objective is to find a way to implement requirements, respecting constraints, and ensure good quality. According to the ISO standard 9241-210 [1], the design process of an interactive systems is iterative: needs and requirements are analyzed, design solutions are identified, created, tested and improved until the development team produces a deployable version of the full-fledged interactive system that meets users' goals and needs. This process produces two types of results: a specification of the design solution to be implemented (the interactive system) and a set of design decisions that drive the evolution of the design along the iteration cycles.

* Authors' addresses: Marco Winckler, marco.winckler@univ-cotedazur.fr, Université Côte d'Azur, SPARKS/wimmics team, CNRS/INRIA I3S, France; Philippe Palanque, palanque@irit.fr, Universite Paul Sabatier - Toulouse III, IRIT, France; barboni@irit.fr, Universite Paul Sabatier - Toulouse III, IRIT, France; Jean-Luc Hack, jean-luc.hak@softeam.fr, Softeam, France; Olivier Nicolas, Softeam, olivier.nicolas@softeam.fr, France; Laurent Gonçalves, laurent.goncalves@softeam.fr, Softeam, France.

We claim that the information explicitly described by models/artefacts are not enough to describe all the information required to produce interactive systems, hence the need of annotations. To illustrate our point, consider the case of user interface prototypes that are the most common type of artefact used to describe design solutions for interactive systems. In all the possible forms (i.e. low-fidelity, hi-fidelity, executable, etc.) prototypes feature a concrete (yet partial) representation of an interactive system and they can be used to explore many design alternatives before implementing the final product [2]. In early phases of the development process, drawings and wire frames are useful and desirable [3] to support ideation of the product but as the process advances, they are replaced by interactive specifications and by executable prototypes [4]. Prototypes are useful and necessary but they are not sufficient to fully describe an interactive system. On one hand, prototypes are not self-explicative, which is illustrated by the fact that annotations are widely used to explain, for instance, the use of icons in a design [5]. On the other hand, prototypes cannot directly inform aspects of the interactive system such as the goals they support (and the ones they don't) and how the system will compute the information provided by users. One might be tempted to assume that lack of expressiveness of prototypes would be solved by using multiple artefacts/models (such as task models, dialog models and interaction technique models) to provide complementary views. Nonetheless, handling multiple artefacts implies that related artefacts should be updated together during the iterations in order to keep a consistent and integrated view of the interactive system [12]. For example, adding a new button to a low fidelity prototype should affect the tasks performed by users and the dialog model describing the inner system behavior. Therefore, artefacts must co-evolve along the development process [10].

Whilst decisions made by the development team occur iteratively throughout the process and will result in evolutions of multiple artefacts, the ISO standard 9241-210 is very silent about how to record and process such design decisions. Empirical observation [14][15] have shown that development teams often make extensive use of annotations. Annotations are flexible, they can assume many forms (such as text, sketching, etc.) and be attached to any type of artefacts, making them a suitable mechanism to provide complementary information throughout the development process. The study performed by Gutierrez et al. [15] pointed out that annotations are used by members of development teams to: i) record the results of discussion including decisions and upcoming tasks, communicate and inform other team members of the work done, ii) gather internal and external feedback on artefacts stored in the workspace, iii) conduct usability evaluations by documenting information and by recording conversation between design teams and UX experts, and justify design choices, and document the design choices by describing them retrospectively. While previous work [15] argued for the use of annotations for user interface prototypes, this paper extends the use of annotations made over various artefacts (including prototypes) by providing a generic engineering solution to connect them across multiple artefacts. More precisely this work presents a design and engineering solution that allows using annotations to achieve the following goals to:

i) Enrich artefacts used to specify an interactive system with any kind of complementary information that would be used to support decision making along the development process;
ii) Cross-reference and cross-check design decisions to multiple artefacts;
iii) Follow the evolution and ensure the consistency of design decisions along the development of interactive systems.

To achieve these goals, we put annotations as first-class information throughout the design process. We propose a model-based approach offering a unified view of annotations throughout the interactive system iterative design process. The meta-model extends the W3C's Web Annotation Data Model by revising concepts (such as authors, versioning and annotation types) to better support interactive systems designers making it possible to track changes made in artefacts and their information. Our approach is generic to any type of artefact but the present work focuses on three types of artefacts: task models, dialog models and prototypes.

Hereafter we start by revising the concept of annotations (section 2) and we present an overview of existing studies about the use of annotations over prototypes. Section 3 review existing annotation models proposes an extension to the W3C's Web Annotation Data Model [16] to support the annotations on multiple artefacts. Section 4 introduces an approach and a tool support (called ARMADILLO) for connecting multiple artefacts using annotations according to annotations model presented in section 3. In order illustrate the feasibility of our approach, we illustrate the use of the tools on a case study on the domain of aircraft cockpits. The last section concludes the paper and introduces perspectives and future work.

## 2 RELATED WORK

In this section we present an overview of the literature concerning the structure and the use of annotation for the design of interactive systems, including tool support for prototyping activities.

### 2.1 Anatomy of annotations

The first studies about annotations started with the identification of common practices by university students on their paper textbooks [17][18][19]. Many of the elements of paper-based annotations were then transposed to electronic documents. Therefore, the common definition of annotation such that provided by Bringay et al. [19] below, refers to documents: "*An annotation is a particular note linked to a target by an anchor. The target can be a collection of documents, a document, a segment of a document (paragraph, group of words, image, part of image, etc.), and another annotation. Each annotation has content, materialized by an inscription. It is the trace of the mental representation elaborated by the annotator about the target. The content of the annotation can be interpreted by another reader. The anchor links the annotation to the target (a line, a surrounded sentence, etc.)*". Looking to more robust definition of annotations for digital documents, we have found annotations describe as "user made statements", consisting in a body (i.e. text note or graphical content), a link (or anchor) to a target including a location within the document as well as other metadata [22]. In the Open Annotation Data model and the Web Annotation Data model proposed by Sanderson et al. [20], [21] annotations are considered as a set of resources in which the body is related to one or several targets (document annotated) as illustrated by Figure 1.

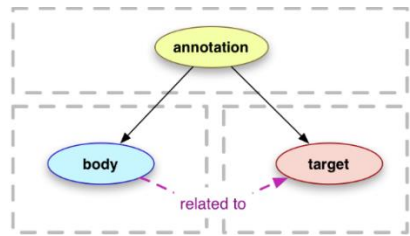

Figure 1 : Simplified model of annotations defined by the Web Annotation Data Model.

All those definitions acknowledge the distinction between an element being in relation to another: the annotation body and the annotated document, which are linked somehow by an anchor. Agosti and N. Ferro [13] stress the importance of the linking mechanism to formally describe annotations. This linking can be made in several ways, depending on the support of each part of the annotation. For example, when annotating books, notes can be written in the margins or in post-its. Marshall [18] identified 3 different mechanisms of association on textbooks including: arrows to connect the document to its annotation, marks such as bracket and brace, and the proximity of the annotation and the target by writing in the margin or the interline. However, as for digital annotations, the artefact annotated and the annotation can be located in the same or in two different files. For that, we talking about electronic documents we add to this list of linking mechanisms the usage of reference, which allows a nonintrusive way to annotate a document. Moreover, the linking between a body and a target created by an anchor might convey particular meaning for the annotations, as suggest by S. Bringay et al. [19]:

- The placement of an annotation can be explicit (when the target is clearly visible in document), or tacit anchor (when placed in the document but not connected to a particular element in the document).
- One-target versus multi-target anchor, with respect to the many possible targets for an annotation;
- Conventional and not conventional anchor, with respect to the existence (or not) of an agreement for interpreting annotations (red marks means might refer to a convent for important topics).

As we shall see, these three elements (body, target and anchor) are core concepts not only for paper-based or electronic documents but they are essential to understand how annotations applies to the many artefacts used to build interactive system as well. In [36], Li et al. have defined a classification of annotation approach for Computer-aided Design. This classification identifies the following categories of attributes that complete the specification of annotation: targeted media, audience, rendering system, usage and function, representation, and storage location. This classification of annotations brings another complementary view of annotations.

## 2.2 Function, meaning, and uses of annotations of digital artefacts

Based on the review of the literature, we summarize hereafter three main functions played by annotations: to enrich a document, to support communication and to support an intention/activity carried out by the author of the annotation. Whilst most of the literature in the matter refers to text documents, we suggest that the following classification is relevant for the development of interactive systems as it can help to add semantics to the widget used for creating annotations over artefacts.

*2.2.1 A Mean to Enrich Documents*

When adding an annotation, the document is augmented somehow. Based on the analysis of the writing contents of the body of an annotation and the relationship created to the target, Zacklad [24] suggests there are three types of annotation:

- An attentional-annotation draws the attention of future readers of the document by emphasizing and pointing out some part of the document. For example, highlighted text, underlined text, symbols.
- An associative-annotation connects an existing element to another one. This annotation can be represented with arrows or references that are not necessarily located in the same document.
- A contributive-annotation is a new information created in reaction or in response of a segment of the annotated document. This annotation either complete this document or discuss it and it requires a link with its initial document by using an associative-annotation. The contribution is either added to the document or in the next edition of the document (which is a similar concept of the elaboration annotations proposed by Lortal et al. [26]).

To this list, we add descriptive-annotation such as semantic annotation as defined in the W3C [21]; which consists in adding metadata to make it easier to process by computer (e.g. indexing, researching) and make them interoperable between different systems. For example, a document is extended with structured information such as the author, the title, the creation date. G. Lortal in [26] qualify those annotations as "Computational level" annotation.

*2.2.2 A Mean to Support Communication*

Annotations play an important role for the communication between diverse actors (author, reader, reviewer, coordinator, etc.). In an iterative design process, the interaction between actors will ultimately make artefacts and annotations themselves to evolve over time. Bringay et al. in [20] defines collaborative annotations as a way to help actors to communicate in a collaborative work to accomplish three main goals: i) Editorial help: annotations can be used as a guide for the creation of a document by indicating instruction or constraints. They can be used as a set of guidelines for the creation of the document or in a revision of the document after a review; ii) Argumentation: annotations can also be used to discuss and argue between collaborators about the document; and iii) Planning: annotations can be used to coordinate the project, plan tasks to do and manage the people working in the project. Three attributes are related to this type of annotation: the task to carry out, the time to complete the task and the person or group of people in charge of the task [27].

*2.2.3 A Mean to Convey the Large Variety of Authors' Intentions*

Another classification of annotations refers to the authors' intention and/or activity carried out by the author of the annotation. In this respect, the classifications proposed by Naghsh et al [24] and by Agosti and Ferro [10] are worthy of mention. Naghsh et al [27] identified 6 different usages of annotations that match with the categories defined above:

- Clarifying and explaining the design.
- Verifying and request a verification from other designers or users.
- Exploring by asking questions to obtain more details on end users' needs.
- Altering or requesting an alteration proposed by the end users.
- Confirming and give feedback on a design.
- Understanding by asking questions to the designers.

Agosti and Ferro [13] encompasses three goals:

- Comprehension and study. The intention here is to understand and analyze the document. It can relate to attentional-annotations in which the annotator highlights parts of the document he found interesting or ask questions to help his comprehension. Those annotations do not add information on the document.
- Interpretation and elucidation. This usage refers to the annotations made to add information, explain a document according to the annotator understanding in order to make it easier to understand and then discuss about it. It could be an analysis or an argumentation for instance.
- Cooperation and revision. Annotations can also be used for sharing ideas and opinions about a text. This can be done through evaluation of the document, feedbacks on it or tasks planning for example.

As we shall see, goals described for annotations might be present in design activities on interactive systems.

**2.3 Use of annotations for the design of interactive systems**

During the design process of an interactive system, the development team gather information about users and produce artefacts to represent the design solutions. The fully-fledge system is the result of iterations of cycle of design an evaluation of the artefact produced. Design decisions are made along the development process and influence the next step of specification of the system. As illustrated Figure 2, annotations are indeed commonplace and might have multiple uses during the development of user interface prototypes. Given the associative nature of annotations, it is quite natural to consider annotation as a possible design solution to the problem of tracing design decisions to artefacts. The annotations presented in Figure 2.c, for example, should be interpreted as a design decision.

Gutierrez et al. [28] conducted a study to investigate how annotations could affect the development of interactive systems in a UCD process. For the purpose of that study, they developed an independent tool called Helaba, which allow the design teams to store and organize artefacts on a common workspace and to connect their decisions to them. The ultimate goal was to support the traceability of design along the design process. Annotations are materialized by Heleba by the means of "Decision cards", "Notes" and "Conversation thread" that can be attached as references to artefacts in order to add content or discuss about an artefact or a specific part of it. They observed that participants stored a curated selection of artefacts that were representative of the process in the shared workspace during the different activities of the UCD and that the annotations provided in her tool were used by the participants of the study to "build a narrative of their design process, especially in relation to how artefacts linked to each other". According to [28], annotations were used to:

- Record the results of discussion including the outcome of those discussion, decisions and upcoming tasks.
- Communicate and inform other team members of the work done.
- Gather internal and external feedback on artefacts stored in the workspace.
- Conduct usability evaluations by documenting information and by recording design team's comments.
- Remind and to justify choices that was made during the process "in the late stages of the project".
- Help to document the design choices by describing them retrospectively.

Overall, this study showed the usage of a shared workspace of annotations that were used along the design process (in particular for User analysis, Task analysis, Lo-Fi prototype and Hi-Fi prototype). Nonetheless, Heleba works as a repository of artefacts and annotations that are not directly connected to the tools used to build the artefacts.

It is interesting to notice that annotations must be considered a special case of artefact required to follow the development process of interactive systems. One particular aspect of the annotations in a UCD process is that annotations can be related to certain versions of the artefact but not each of its version (e.g. an annotation indicating to fix an error). Moreover, annotations (especially contributive or organizational annotations) can influence the evolution of other artefacts but also affect by the evolution of the artefacts. Thus, while annotation have their own lifecycle, this lifecycle can be affected by the related artefact's lifecycle as well.

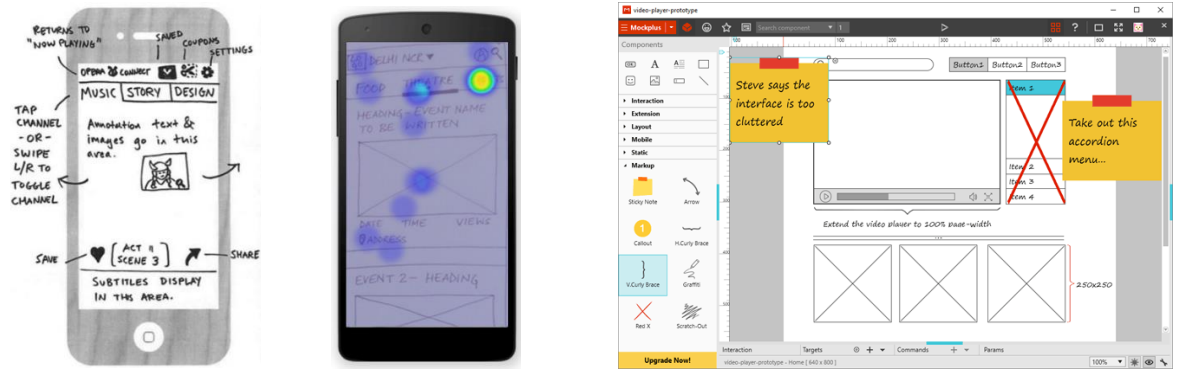

a) Explaining the design.    b) Record results of usability evaluation.                    c) Document design decisions.

Figure 2 : Three examples of the use of annotations on prototypes.

## 2.4 Tools for supporting annotations of prototypes and other artefacts

In this section we present a summary of an analysis of 80 tools for annotating prototypes. Using the keyword "annotations tools", "prototyping tools", and "wireframing tools", we have retrieved publications from conferences (including CHI, UIST, DIS, EICS, INTERACT and IAnnotate) and then we extended the search to find other tools indexed by google.com. These tools can be generally classified into two categories: a) generic annotation tools that are loosely coupled with prototypes and artefacts used for the design of interactive systems, and b) prototyping tools that embed annotations as native functions for designing interactive systems.

Among the 25 generic annotation tools, we have 18 academic tools which means tools described in scientific publications (i.e. *Amaya, Quilt, sense.us, HyperImage 3, SparTag.us, D.note, LiquidText, List-it, Zydeco, Annotorious, ChronoViz, Domeo, Elias' prototype, GatherReader, Instant Annotation, Neonion, Dokieli,* and *Pundit Annotator*) and 7 commercial tools (i.e. *Annozilla (Annotea project), Diigo, Protonotes, Annotatorjs, Authorea, Hypothesis, Ponga*). The complet references to these tools are available in the Annex I. Among the prototyping tools embeding functions for annotating prototypes, we have found 55 tools where 3 are academic tools (i.e. Silk, DEMAIS, ActiveStory Enhanced) for the other 52 have been distributed as commercial tools at some point (i.e. *Rise, SoftAndGui, UxPin, AdobeXD, Microsoft visio, Smartdraw, Axure, GUI Design Studio, MockupScreens, JustinMind, Balsamiq, DesignerVista, InPreso Screens, MockingBird, PencilProject, Pidoco, ProtoShare, WireframeSketcher, Cacoo, Crank Storyboard Designer, Creately, FlairBuilder, ForeUI, Gliffy, Microsoft SketchFlow, iPlotz, BluePrint, FrameBox, HotGloo, LucidChart, Mockflow, Sketch, Antetype, Draw.io, Lumzy, MockupBuilder, Mockups.me, MockupTiger, PowerMockup, Proto.io, FluidUI, IndigoStudio, Moqups, Prototyping on Paper (Marvel), Alouka, Concept.Ly, InVision, NinjaMock, Notism, MockupPlus, SnapUp,* and *Atomic*).

A full comparative analysis of these tools is to be found in [29]. Hereafter, we provide a summary of the most striking findings that we consider relevant for understanding the contribution presented in this paper.

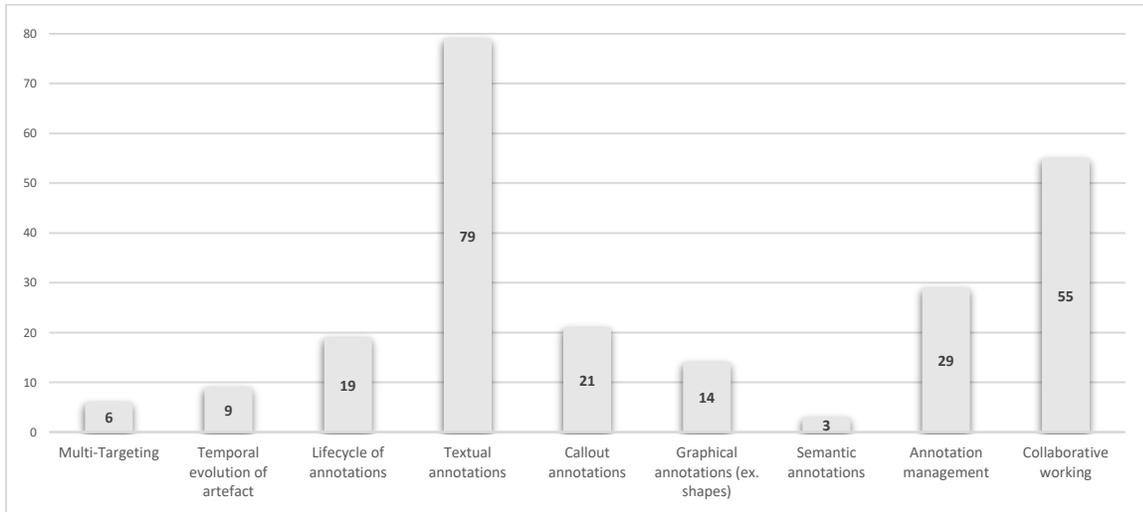

Figure 3 : Number of tools supporting nine different features concerning annotations (i.e. multi-targeting, temporal evolution of artefacts, lifecycle of annotations, textual annotations, callout annotations, graphical annotations, semantic annotations, annotations management, collaborative word). in a comparative study of 80 annotations tools.

During our review, we observed that annotation support in the tools were limited for our studies. Indeed, we did not find a satisfying tool providing support for our needs on evolving artefacts and annotations management in a UCD process. The main objective of the annotating tools we reviewed is to provide a medium that can be used to communicate over a document. Most of the annotating tools only allow to annotate web pages. As for prototyping tools, annotations features were either promoted for helping collaboration within the design team or to gather feedback from external users. The overall result of this study is that while annotations are supported in many tools, their implementations are limited and not always suited in the context of the iterative process of a UCD approach.

Hereafter we list the main findings of this review:

- Annotations can only be related to one artefact at a time. Other artefacts can only be loosely referenced by citing them or enclosing a copy. Thus, creating a unidirectional relationship between those artefacts. Only 6 tools allowed to set several targets on one annotation.
- The targeting of annotation is either made by selecting text, defining areas, positioning the annotation, positioning a marker or citing the related fragment of artefact.
- Temporal evolution of artefacts is rarely acknowledged and it is managed by only 9 tools.
- Fewer tools manage to keep a consistent link between the annotation and its artefact either by extracting a snapshot of the targeted fragment or by implementing a limited recognition system.
- Lifecycle of annotations has been considered in only 19 tools through a status system allowing users to manage manually annotations. The other tools forces user to dispose manually the annotations once they have been processed which imply a loss in the traceability of the management of annotations.
- Textual annotations are the most adopted form of annotation followed by graphical representation of either shapes, icons or markers. Only three tools enable to enclose files in annotations (JustInMind , MockFlow and Notism).
- There is no integrated support for documenting a custom semantic. Thus, those semantics are either implicit inside the design team or defined in an external document. This semantic can be used for managing the annotations or give them a weight. When available, semantic is limited to the use of predefined tags (e.g. Amaya , iRise , Concept.Ly , Quilt [30], Cacoo, Neonion [31]).
- As for the management of annotations, 29 tools feature a list of the annotations. Following this list, we noted that only 17 tools supported a filtering option and 15 tools supported a search feature on the annotation. A navigation feature between annotations and their targets is not systematic in every tool we reviewed.
- For the collaboration, 55 tools supported a synchronous collaboration over artefacts. In 27 tools, it is possible to restrict users' right on the artefact or in the annotations. In 42 tools, author of an annotation can be identified.

### 2.5 Pitfalls of existing annotations approaches

This study about annotations raises some interesting questions about the role of annotations for engineering interactive systems. It is worthy of notice that annotation is a concept easy to understand and common place in many professional scenarios where a piece of information cannot be represented as being part or the artefact. Empirical studies demonstrate the important role of annotations in communication between teams and support for decision making processes along the development process of interactive system; and yet quite often annotations are still considered as a second-class piece of information that is not systematically treated (or even mentioned) by most development approaches.

Annotations can have different formats and be used for different purposes with multiple artefacts. Nonetheless, existing tools only support a few types of annotation formats (mainly text and drawings) that lack of semantics to support decision making process.

Annotation tools are often very specialized for a single type of artefact. Moreover, there is a strong binding between the annotation tool and the editor. Whilst such strong binding make sense for an interaction point of view, it is not possible to reuse annotations outside the editor which a major drawback for tracking annotation created on multiple artefacts.

Development processes using multiple artefacts are quite common and yet, they fail to provide a uniform view for annotations that could make sense multiple artefacts at a time. It is interesting to notice that previous work [51] pointed similar problems when dealing with multiple UML models; however, we could not find in the literature a solution for unified view on annotations.

Conversely to other approach, the solutions we propose were designed to support systematic use annotations of multiple artefacts along the development process. First of all, our solution allows to connect multiple artefacts (as illustrated in section 5). We rely on standards (described in section 3) which allows the export/import of annotations into diverse tools. We differ from other approaches where the annotation mechanisms are a functionality of the editor of artefacts (strong binding between artefacts and annotations), by proposing a distributed architecture (section 4) where annotations created using plugins are stored independently into a central storage, thus keeping annotations available for reuse by other tools (including the creation of dedicated tools for creating an overview of all annotations in different artefacts of a project). The architecture of plugin is meant to extend editors of artefacts. Last but not least, as part of our work we have proposed (see section 4) a set of annotations formats that go beyond text or drawings (including markers, voting mechanisms, and testing scenarios) that features the semantics for the use of annotations.

## 3 ANNOTATION MODEL

Currently, the Web Annotation Data Model proposed by the W3C [16] is the standard model for describing annotations. The W3C annotation model provides a common format for describing interoperability of annotations through the web. In this section, we propose an extension of the Web Annotation Data Model [16] to cope with the idiosyncrasies of annotations of interactive system design and development.

### 3.1 Rationale for extending the Web Annotation Model

The Web Annotation model [16] was initially designed to support the annotation of Web documents and URIs (Universal Resources Identifiers). As such, the Web annotation model is quite generic for annotation of image and text documents. That model is quite flexible and meant to be extended according to the particular cases of use [21]. By extending the Web Annotation Model, we aim at reaching the following idiosyncrasies of interactive system design and development:

- *Annotations concern diverse design artefacts:* during the context of the design of an interactive system, a large variety of artefacts is produced. Annotations should work with all sort of artefacts required for the design and development of interactive systems. In the present work, we are particularly interested in artefacts including prototypes, tasks models, system models, specification documents and any other document produced or gathered during the design process. Each artefact is likely to receive annotations either as a support for active reading, for communicating, for reviewing, for planning or for editing.
- *Independence with respect to the targets:* annotations are often considered part of the artefact and/or strongly tied to them. Such as a strong binding prevents reuse and analysis of annotations outside the environment where they were created. Some level of independence from the target might allow the export of annotations, so that they can be used as a first-class information to reason about comments and decisions made during the development process.
- *Connecting multiple targets:* an annotation might concern different artefacts. For example, a typo in a form field not only affect the user interface prototype but also the data model. Enabling the connection to multiple targets helps to reduce redundancy of annotations referring to the same problem and also foster the connection between artefacts that are implicitly related.
- *Support the evolution of artefacts:* artefacts are produced through iterations and continuously evolve along the development process. The development team might need to trace of the outcomes of annotations attached to evolving versions of artefacts.
- *Consider the life cycle of annotations:* along the development process, annotations might be *created, revised* by the creator and/or other members of the development team, and *disposed* when no longer needed. For that, annotation must be part considered as living components that can change the status according to a specific life cycle that might be independent of the evolution of targets (the design artefacts).
- *Enabling many forms of annotations:* annotations might be created using diverse forms (such as text, brushing/highlight markers, drawings, icons, …). They can superimpose contents available in other sources, for example adding a layer information about usability problems found on design artefacts [11]. Complex annotations forms might include voting mechanisms as a mean to consolidate opinions around design options [14]. Each form of annotation might be more or less suitable to a particular type artefact. Nonetheless, each form of annotation must have a specialized selector that allow users to express an intention with respect to artefacts. Choosing a particular form of annotation (ex. text or drawing) might also imply a particular meaning and require a particular processing.
- *Make explicit the intentions:* annotations can be adapted to many usages and functions like communication, planning, contribution in the edition of the artefact and so on. We take the benefice of the possibility of extending the classes of the model to implement different type of annotations (i.e. attentional, associative, and contributive annotations) as suggested by Zacklad [24]. The association of intentions to annotations is a means to make orient decisions along the development process. It is interesting to notice that the intentions for creating annotations might be different according the people involved in the development process. For example, end-users, clients (or product owners), as well other members of the development team might have a different perspective for the usage of the system [46]. So, knowing the profile of the creator of annotations is important for understanding the intentions.
- *Add semantics to the annotations:* The integration of semantic annotations will help organizing information, building a structure for the information gathered and describing the interactive system being developed in its entirety instead of only considering localized annotations in artefacts as existing prototyping do tools do. The semantic annotations describing the relationship between artefacts, annotations for planning according to Bringay [19] as well as annotations for documenting the design process that are described by Gutierrez et al [15]. Our extended model aims to support both semantic annotations and non-semantic annotation on artefacts in the context of the design process of an interactive system. It is compatible with the attribute categories defined by Li et al. in [36].

These specificities of the design process of an interactive system indicates a particular use of annotation that goes beyond the usual annotation of documents. Indeed, we need to consider the evolution of the artefacts, the lifecycle of the annotations, and the evolution of the information on the interactive system being designed (which might include the successive versions of the artefact as well as the input provided by every actors of the design process).

## 3.2 Macroscopic view of the annotations model

Our annotation model consider two types of classes of particular importance: the classes allowing to define the characteristics of annotation (which includes the type of annotation, the description of the intended audience for the annotation completed by the motivation and the purpose of the annotation stating the reasons of the creation of the annotation) and the classes describing annotation metadata (including information about date of creation and modification of an annotation and identity of its creator).

We take the benefice of the possibility of extending the classes of the Web Annotation Model to implement different type of annotations. The core classes of our extended model are: **annotation, target**, **creator** and **artefact**. Notice that the first three classes are already present in the Web Annotation Model, while the class **artefact** is a new addition.

Figure 4 provides a view at glance of our model, highlighting: i) classes defined in the W3C model (in orange), ii) new attributes that have been added to enrich metadata classes (in violet), iii) new classes proposed as an extension to the model (in blue), and iv) specialized classes that extend the use of the overall model for particular types of annotations used in our case study (in green).

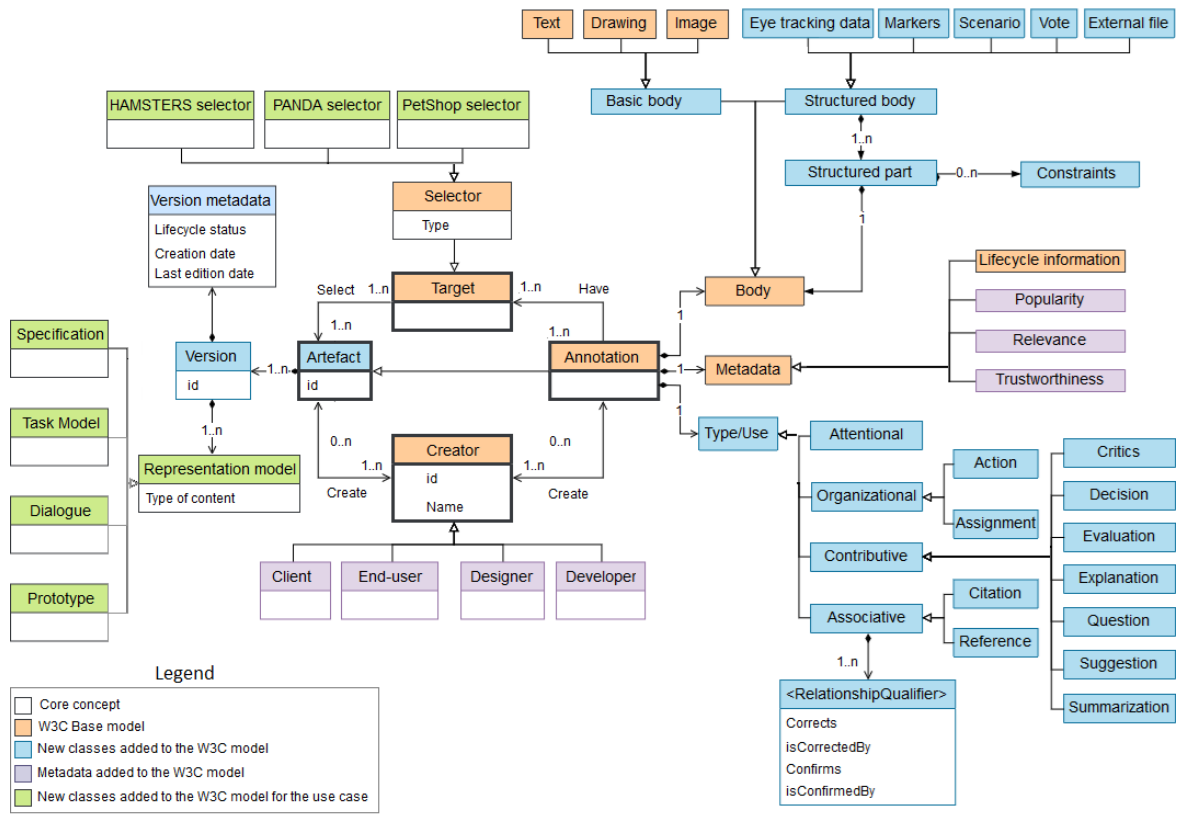

Figure 4 : Overview of extensions made to the W3C's annotation model.

The class **artefact** was added to the model to allow adapting an annotation to the diverse types of artifacts that might be used along the development of interactive systems. Is considered an **artefact** any file (or document) that is produced (or gathered) during the design process. It is composed by a set of versions that correspond to a versioned content of the artefact. Each version is also characterized by its metadata which contains various information like the creation date and the lifecycle status of this version that indicate if the artefact is being written, waiting for review, reviewed finished, being updated or archived. The content of the artefact refers to the type of information that is contained in the artefact. Each type of content reflects a different aspect of the interactive system that is being designed and each follow a different syntax. For instance, a task model can be described using a task model notation such as HAMSTERS [41].

In our model, an **annotation** considered is a particular type of artefact. This enable the independence of annotations compared to their evolving targets, it enables annotating different artefacts with one annotation and also enable annotating other annotations. Moreover, our model extends the elements *body*, *type* and *target* as defined in the Web Annotation Data Model. As for *body*, our model includes classes allowing adapting the visual aspect of the annotation. We consider a body might have a basic representation (such as text, drawing and images) or a structured representation composed of (at least) a label and an interactive element. Examples of annotations featuring structured body include: a vote (which contains a counter allowing every reader to cast a vote agree/disagree), a scenario (which features a list of text elements constrained by a grammar to express a sequence of tasks performed by users), an external file (which in addition to a visible label, has a link allowing to open the file), and markers (associating a glyph/icon next to the label). As for the *type*, this was extended to describe the function of the annotation as suggested by Zacklad [24] (i.e. attentional, organizational, contributive and associative annotations).

Like the Web Annotation Data model, a **target** is materialized by a **selector** which is used to specify the relevant parts of the artefact that is being annotated. A selector must be specialized to cope with the inner structure of the document being annotated. These extensions describe how the model can be adapted (using an adapter design pattern) to cope with the specificities of every development environment for editing the diverse artefacts used for building interactive systems. So far, we have extended this class to copy with HAMSTERS task models [41], PETSHOP dialog models [45] and PANDA prototype models [14]. We illustrate here how specialized selector are created in the model. Further details about these models are presented in the case study.

We also extend the entity **creator** defined in the W3C which was originally meant to identify the people or group of people that create the annotation. Indeed, we refine the entity creator to include a variety of roles involved in the development process of interactive systems. In our model, these roles are expressed by the means of metadata whose terms must be adapted according to the project. A basic list of roles having access to the annotations includes the *client* (or product owner), the *end-users*, the *designer*, and the *developers* (this should encompass the role of participants involved).

Metadata is used to assess the different annotations produced on the artefacts. Annotations can be created by any user and each annotation can express a variety of information on any domain (ex. Use cases, requirements, constraints, data, identification of problems, personal feedback, and personal opinion). The role of the creator of the annotation can help to determine both the expertise and legitimacy of the creator in the information given in his annotations. Other parameters can quantify (popularity of an opinion within an identified group) or be informally assessed such as the relevance and trustworthiness of information (a creator can emit hypothesis or facts based on unreliable sources).

## 4 TOOL SUPPORT: ARMADILLO AND PLUGINS

In order to demonstrate the feasibility of our approach we have developed a tool suite called ARMADILLO, which stands for "*Annotating by Referencing Models, Artefacts, Documents to Identify Logically Linked Objects*". ARMADILLO implements the extended annotation model described in section 3. The ARMADILLO tool suite was specifically designed to help the development team to accomplish two main tasks:

i) annotate artefacts used to specify interactive systems; and,

ii) allow cross-referencing of annotations and multiple artefacts.

To support these tasks, ARMADILLO encompass two main components: a *Project repository* and a *plugin*. The *project repository* is used to manage all the annotations created on a project and deal with the cross-referencing of annotations created over diverse artefacts. The *plugin* allows to create annotations using the editor of artefacts used in the project. All these components of ARMADILLO were built as a Java application over the NetBeans Platform framework to facilitate the integration and distribution of the tools

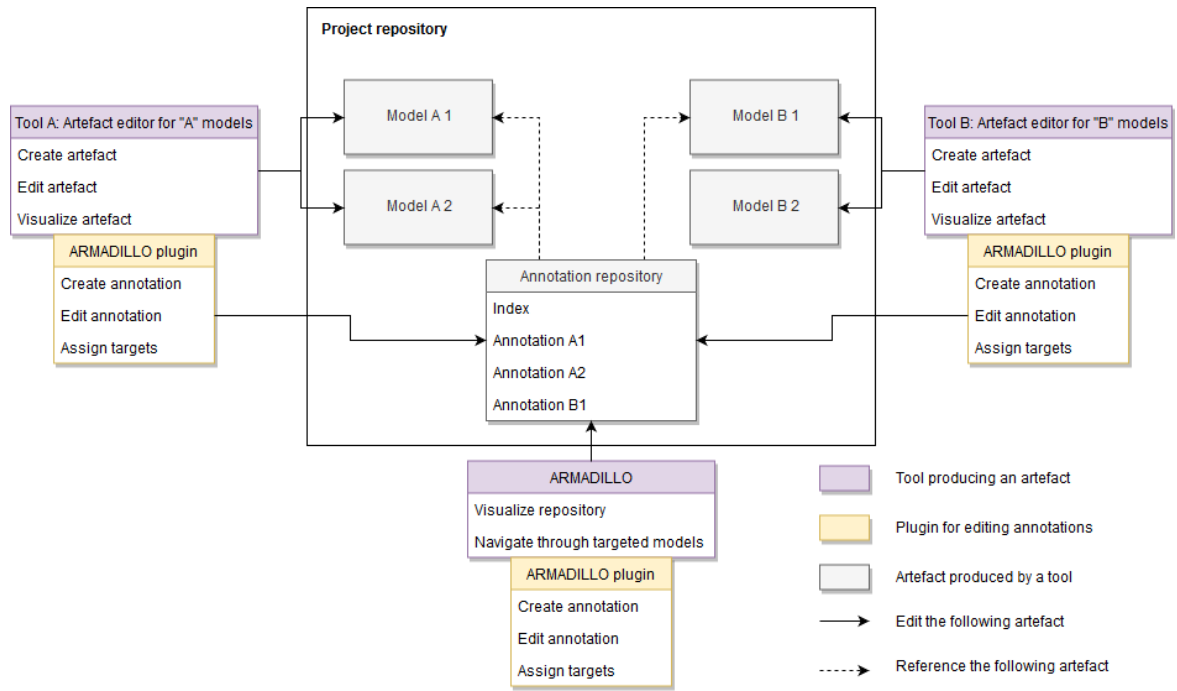

Figure 5 : Overview of ARMADILLO architecture and its components *project repository* and *plugin*.

### 4.1 ARMADILLO: *project repository*

The component *project repository* is the central element in the architecture of ARMADILLO. It was conceived as a central storage of files that contains the specification of the interactive system (i.e. the models as we follow a model-based approach) and all the annotations created over these artefacts. Annotations are stored as independent files in a central annotation repository. The central part of the Figure 5 shows the ARMADILLO repository which contains the list of models describing the artefacts (using generic names such as *Model A1, Model A2, Model B1* and *Model B2*), the list of individual annotation files (one per artefact and per version of the artefact created, respectively *Annotation A1, Annotation A2,* and *Annotation B2*) and an *index* file (regrouping all files in a project). Each model corresponds to an artefact used to specific interactive systems (such as task models, user interface prototypes, dialog models, etc.).

Annotations are stored in ARMADILLO as XML files. Each annotation file contains references for the artefact it annotates, so that it is possible to navigate from a particular annotation to the corresponding the artefact. This feature is illustrated by Figure 6 where we can see an annotation file featuring a reference (see the arrow showing the place) to "*pandaannotation*", which is known in index files records as the artefact editor for PANDA models. With that annotation then, it is possible to know how to launch the corresponding editor (PANDA in this example) to see the annotations in the context where artefacts were built. Other fields in the annotation field (not displayed here) let us to know which project the annotation belong. This basic file-oriented system architecture allows to create annotations at any time along the development process. Moreover, it is possible to connect an annotation to any specification available in the development environment as far as it refers to a file available in the project repository; this is an important aspect allowing the scalability of the approach.

Figure 6 : Structure of an annotation file showing the references to with the plugin *pandaannotation*.

When an annotation is created with ARMADILLO, it is recorded in the *index* file. ARMADILLO uses the index file to parse individual annotations files and generate a graph depicting annotations in a project. Figure 7 illustrates a particular case for an annotation in ARMADILLO that connects multiple targets (see Figure 7.c).

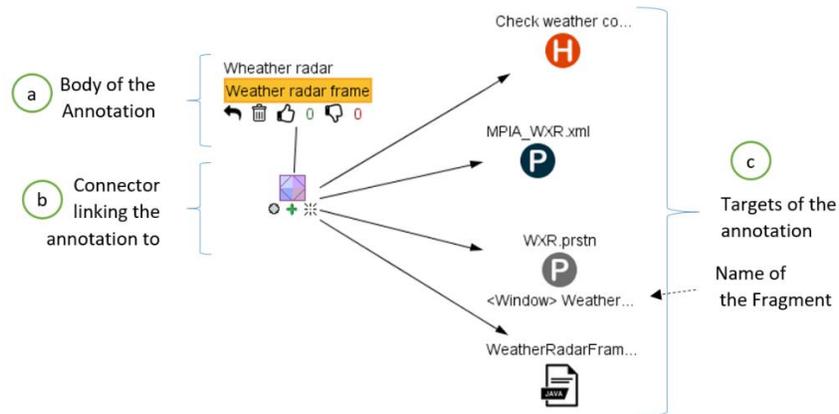

Figure 7 : Example of an annotations in ARMADILLO project repository and plugin.

## 4.2 ARMADILLO: plugins

When annotations are created directly over the repository, the level of granularity of the target is a file that describes the artefacts. For a fine selection of elements (ex. zone, objects, etc.) it is necessary to have access to the editor tool used to create artefacts. For that, in addition to the *project repository*, we deliver a *plugin* allowing to create, edit and assign annotations to targets. So far, the ARMADILLO *plugin* is only available for the NetBeans Platform framework, which means that editor tools build with that same framework. Once connected to the editor, the ARMADILLO *plugin* provides access to a palette of annotations that includes: basic types of annotation (such as *textual_annotation* used for creating text annotations, *drawing_annotation* for hand free sketching) and structured annotations (such as *scenario* for describing a list of steps to follow with the artefact, *annotation marks* that allow to pinpoint elements that might require user attention, and *references* allowing to complete the description with an external file). Figure 8 shows the palette of annotations supported and made easily available through the ARMADILLO *plugin*.

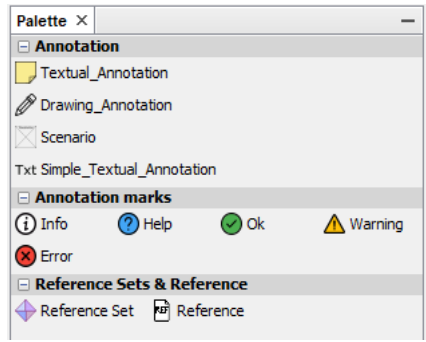

Figure 8 : Palette of the ARMADILLO *plugin* featuring the diverse types of annotations supported by the tool.

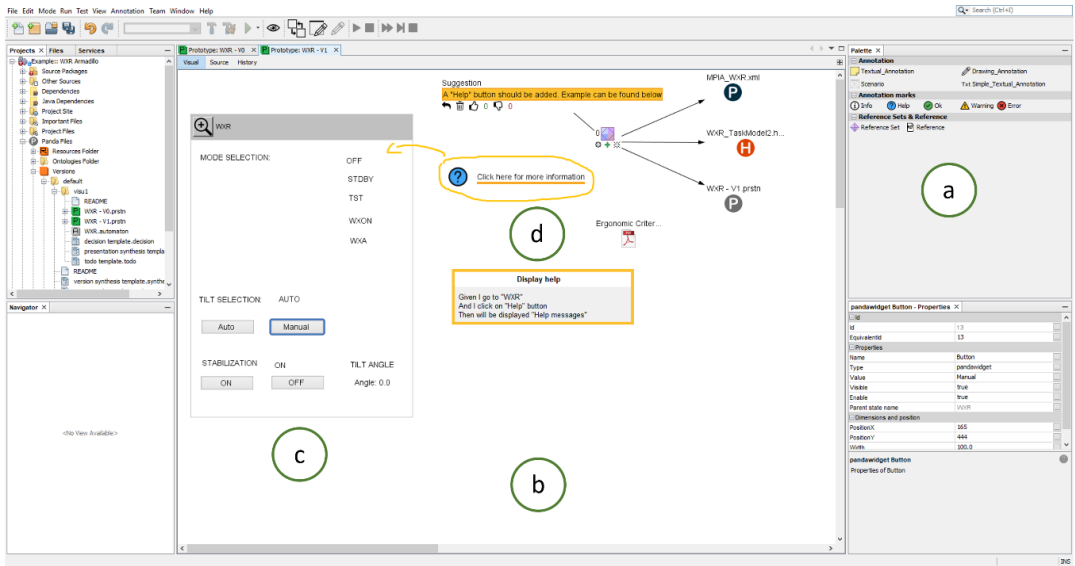

Figure 9 : ARMADILLO *plugin* in context of use as integrated to the tool editor PANDA: (a) palette featuring diverse types of annotations, (b) edition area, (c) prototype artefact, (d) annotations (marks in yellow).

Each plugin is a particular instance of the class *selector* in the model shown Figure 4. Figure 9 illustrates the instantiation of the ARMADILLO plugin in the environment of the editor PANDA. The creation of annotations is then made by simple drag & drop interaction from the palette (Figure 9.a) to the main editor area (Figure 9.b), placed next to the artefacts (Figure 9.c) where annotations (featuring yellow marks, Figure 9.d) can be customized and connected to specific targets.

So far we have developed ARMADILLO plugins for three editors: PANDA (for creating user interface prototypes), HAMSTERS (for describing task models), and PETSHOP (for describing dialog models). These three editors are delivered as part of the framework called CIRCUS [40]; they can be used alone or in combination. All these tools were built upon the Netbeans Platform framework. Hereafter we present a brief description of each tool used for the case study in section 5.

PANDA is a prototyping tool which allow to create medium-fidelity and interactive prototypes by specifying both the dialog and the presentation of an interactive system [14]. This tool support focuses on the modeling of prototypes to formally represent the interactive system as shown by Figure 12. PANDA allows to model the presentation (i.e. screens) and part of the dialog (i.e. navigation between screens) of the interactive system using automata. PANDA prototypes are partial representation of interactive system, thus, they convey a representative view of the current goal of the design process. This goal can be refined or altered during the course of the design process. We suggest that tracing the evolution of this artefact and the reasons of the evolutions would be feasible by structuring the different versions produced within the project workspace of the prototype and by using annotations to connect those evolutions of the prototypes with the other artefacts of the project.

HAMSTERS (Human – centered Assessment and Modelling to Support Task Engineering for Resilient Systems) [41] is a tool-supported task modelling notation for representing human activities in a hierarchical and structured way. Task models [6] are useful artefacts supporting the analysis of user tasks and the description of the logical activities that have to be carried out in order to reach the user's goals. HAMSTERS notation is inspired by existing notations, especially CTT [48], but it has largely extended including pre-conditions associated to task executions, data flow across task models, more detailed interactive tasks… HAMSTERS models can be edited and simulated in an environment which also provides a dedicated API for observing editing and simulation events making it possible to connect task models to system models (such as ICO models). Figure 11 shows an illustration of task models used in the case study.

PETSHOP is a modeling environment that supports the edition, execution and verification of ICO models [40]. The ICO formalism is a formal description technique dedicated to the specification of the dialog part of interactive systems. Dialog models [7] play a major role on design of interactive system by capturing the dynamic aspects of the user interaction with the system providing a specification of the relationship between presentation units (e.g. transitions between windows) as well as widgets (e.g. activate/deactivate buttons). As for interaction techniques models, they enable to describe precisely the events chain (i.e. fusion/fission of events) both at input and output levels [8], thus mapping events to actions according to predefined constraints enabling/disabling actions at runtime which is not covered by other type of artefacts [9]. ICO uses concepts borrowed from the object-oriented approach (dynamic instantiation, classification, encapsulation, inheritance, client/server relationship) to describe the structural or static aspects of systems, and uses high-level Petri nets to describe their dynamic or behavioral aspects. The ICO notation has evolved since to deal with the modelling of multimodal issues in interactive-system (e.g. event-based communication, temporal modelling and structuring mechanism based on transducers in order to deal with low level and higher lever events) and to address news challenges raised by the various application domains it has been applied to, for example VR systems [37] and cockpits in aircrafts [40]. Figure 13 is an illustration of ICO models used in the case study. ICO models can be executed along the application, making this a powerful tool for analysis of the inner system behavior of interactive systems.

## 5 CASE STUDY

This section illustrates the use of the tool ARMADILLO in the context of a real-life case study. Only a partial view of the project using the annotation is given hereafter. The evolution of the artefacts following the production of annotations is not presented here.

The application selected for a case study is a Weather radar (WXR) currently deployed in many cockpits of commercial aircrafts, as illustrated at Figure 10. It provides support to pilots' activities by increasing their awareness of meteorological phenomena during the flight journey, allowing them to determine if they may have to request a trajectory change, in order to avoid storms or precipitations for example. Figure 10.a) presents a screenshot of the weather radar control panel, used to operate the weather radar application. This panel provides two functionalities to the crew. The first one is dedicated to the mode selection of weather radar and provides information about status of the radar, in order to ensure that the weather radar can be set up correctly. The operation of changing from one mode to another can be performed in the upper part of the panel (named MODE SELECTION). The second functionality, available in the lower part of the window, is dedicated to the adjustment of the weather radar orientation (Tilt angle). This can be done in an automatic way or manually (Auto/manual buttons). Additionally, a stabilization function aims to keep the radar beam stable even in case of turbulences. Figure 10.b shows a screenshot of the weather radar display produced by the weather radar aircraft system. Spots in the middle of the images show the current position, importance and size of the clouds (importance is capture by color coding).

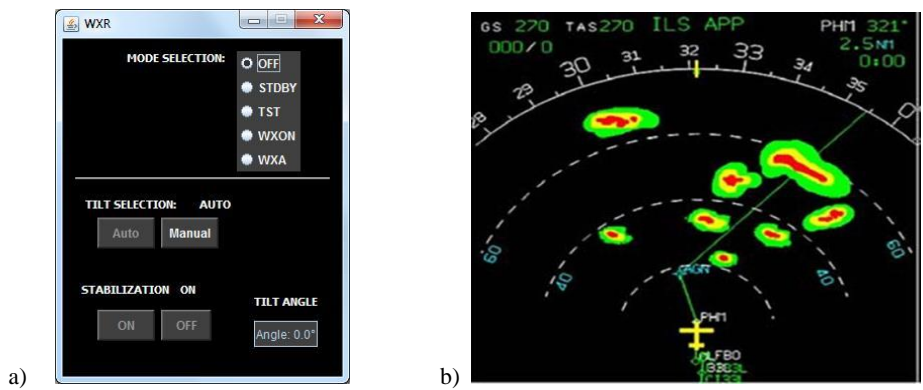

Figure 10 : Screenshots of the case study application Weather radar (WXR). At left (a) the weather radar control pane with the interactive radio button controls. At right (b) the display produced by the weather radar aircraft system.

### 5.1 Various Artefacts of WXR Project

Hereafter we present the various artefacts used during the project of the Weather Radar application (WXR). At this point, the artefacts do feature annotations. Each artefact was selected to provide a unique and complimentary view of the application:

- Task models using HAMSTERS [41] describe the different tasks of the pilot [10] as well as their goals while using the Weather Radar Application (as shown by Figure 11);
- System models using the ICO notation [42] formally describe the system behavior (see Figure 13);
- A medium-fidelity prototype of the user interface designed with PANDA [14] (as shown in Figure 12).
- A High-fidelity prototype of the panel implemented in Java (as shown in Figure 9).

The task model presented at Figure 11 describes the crew activities performed in order to check weather conditions. At the higher level of the tree, there is an iterative activity (circular arrow symbol) to "detect weather targets" that is interrupted (operator [>) by a cognitive task "mental model of current weather map is built". Other human tasks include perception (task "Perceive image") and motor (task "Turn knob"). Connection between crew's activities and cockpit functions is made through interactive tasks (as input "Turn knob" and output "Rendering of radar information"). The time required for performing the latter heavily depends on the radar type. Such behavioral aspects of systems can be modeled using ICO notation and PETSHOP tool as detailed in section below. The task "Manage WXR" is a subtask of a more global activity of flying crew understanding weather conditions ahead of the aircraft. This activity would include analyzing the image produced by the weather radar (shown in Figure 9 b). This task model corresponds to the manipulation of the user interface presented in Figure 9 a. From these models we can see that the tasks to be performed in order to check weather conditions in a given direction are rather complex. The time required to perform them depends on 3 elements: the operator's performance in terms of motor movements, perception and cognitive processing.

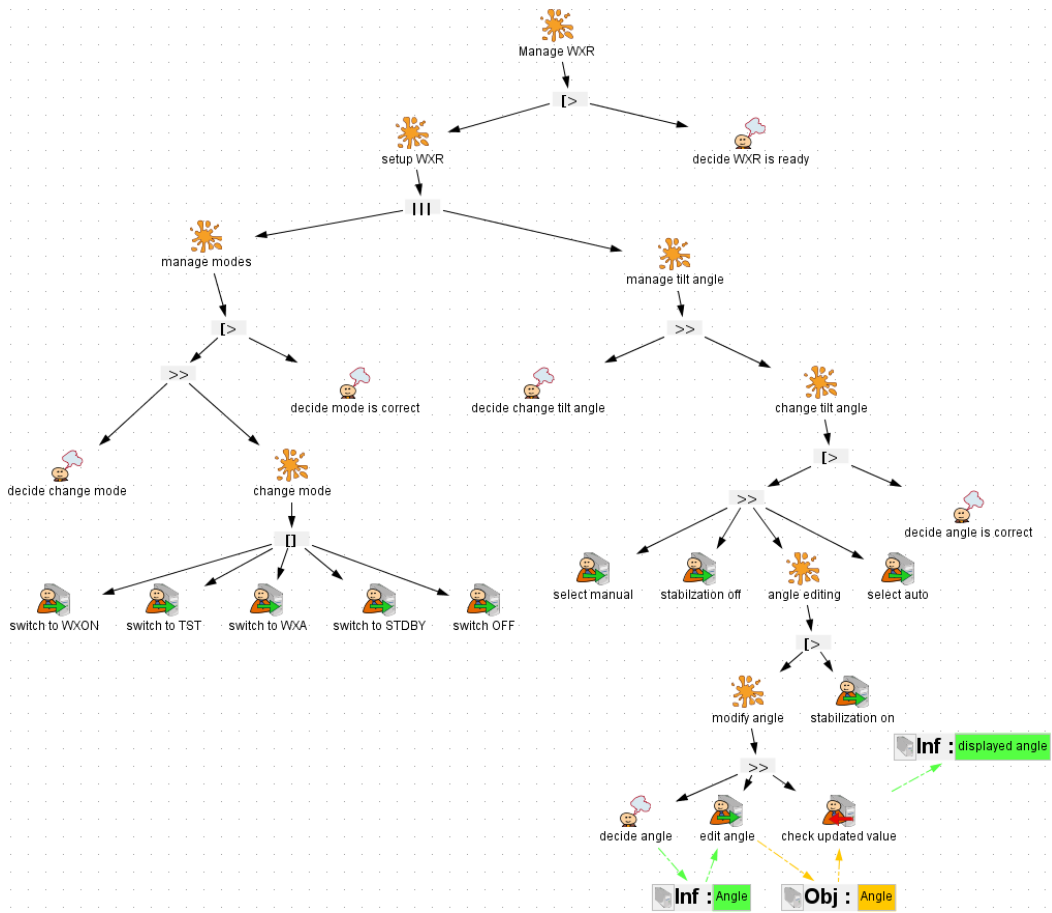

Figure 11 : Task model using the notation HAMSTER describing the WXR project. Hierarchical view of user tasks supported by the application. Different icons represent different types of tasks.

Figure 12 shows the user interface prototype that was designed to provide to the user to:
- Switch between the five available modes (upper part of the figure) using radio buttons (the five modes being WXON to activate the weather radar detection, OFF to switch it off, TST to trigger a hardware checkup, STDBY to switch it on for test only and WXA to focus detection on alerts).
- Select the tilt angle control mode (lower part of the figure) amongst three modes (fully automatic, manual with automatic stabilization and manual selection of the tilt angle).

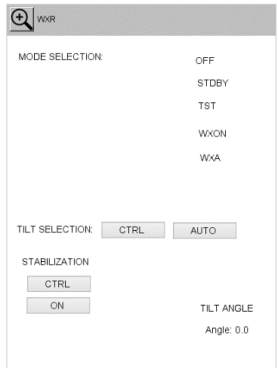

Figure 12 : Prototype created with PANDA for describing the user interface of the WXR project. The screenshot corresponds to an early version of the prototype that evolved along the development process.

The ICO model presented in Figure 13 describes to configure the weather radar using the mode and angle. In Figure 13 presents the behavior description of this part of the interactive cockpit using the ICO model into two parts:
- The Petri net in the upper part handles events received from the 5 radio buttons. The current selection (an integer value from 1 to 5) is carried by the token stored in MODE_SELECTION place and corresponds to one the possible radio buttons (OFF, STDBY, TST, WXON, WXA). The token is modified by the transitions (new_ms = 3 for instance) using variables on the incoming and outgoing arcs as formal parameters of the transitions. Each time the mode value is changed, the equipment part (represented by the variable wxr within the token) is set up accordingly.
- The Petri net in the lower part handles events from the four buttons and the text field (modify tilt angle). Interacting with these buttons changes the state of the application. In the current state, this part of the application is in the state fully automatic (a token is in AUTO place). To reach the state where the text field is available for the angle modification, it is necessary to bring the token to the place STABILIZATION_OFF by successively fire the two transitions switchManual_T1 and switchStabOff_T1 (by using the two buttons MANUAL and OFF represented in Figure 12, making transition change_Angle_T1 available. The selected angle must belong to the correct range (-15 to 15), controlled by the three transitions angleIsLow, angleIsCorrect and angleIsHigh. When checked, the wxr equipment tilt angle is modified, represented by the method called wxr.setTiltangle.

Figure 12 presents the UI of the WXR produced with PANDA, which was designed to evaluate the organization of widgets and information to be displayed in the panel. This prototype is composed of one window called "WXR" and features several labels and buttons. The upper part of the prototype lists the different mode that can be selected: OFF, STDBY, TST, WXON, and WXA. The lower part is used to change the stabilization and the tilt of the radar as well as display the angle of the radar. It is important to note that, with respect to the application in Figure 9.a some user interface elements are not identified yet e.g. the radio button group on the upper part of the window.

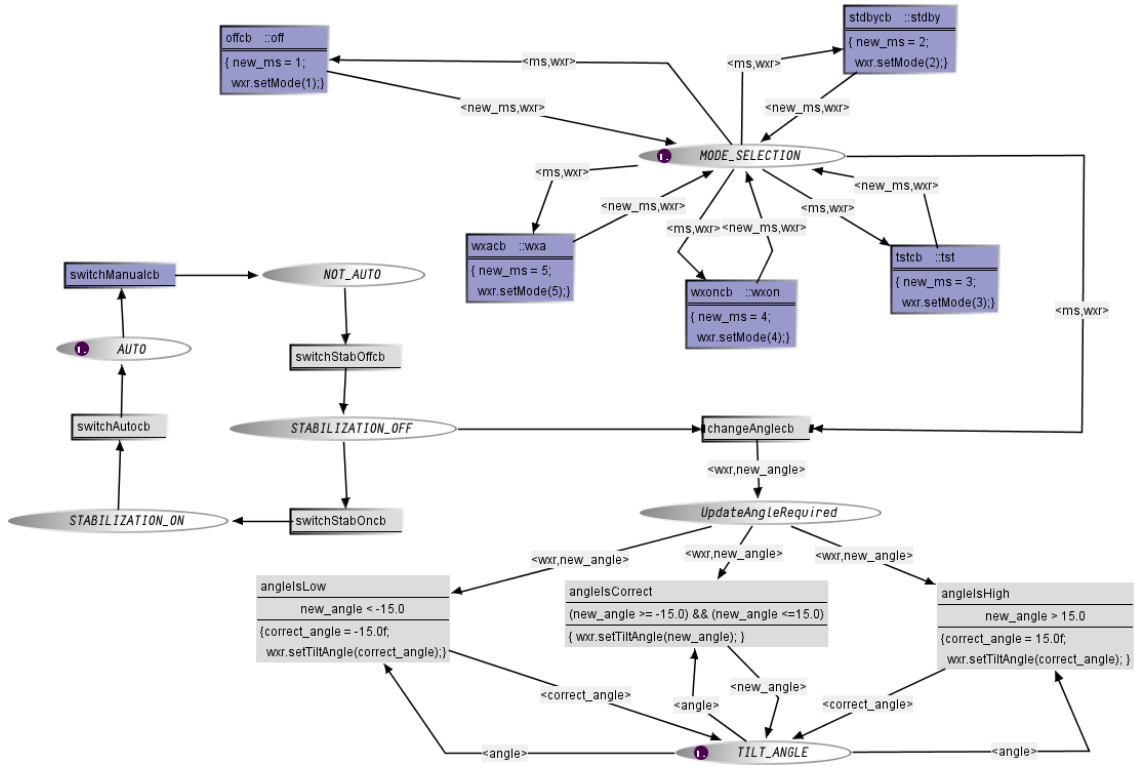

Figure 13 : The system model artefact of the WXR project using the ICO formalism within PETSHOP Tool.

## 5.2  Annotating individual artefacts

Once installed, the ARMADILLO *plugin* features the same on all hosting editors. A palette for annotations (as shown at Figure 8) is embedded into the editor (as shown by Figure 9) and from there the user can annotate individual artefacts. The steps for creating an annotation are illustrated by Figure 14 using the tool PANDA. First of all, we should notice the ARMADILLO *plugin* is installed in the editor (Figure 14.a, at the right-side) proposing a palette with functionalities for creating the annotations. Annotations are created via a drag & drop from the palette to the main edition area (Figure 14.b). From there, the user can select the annotation and type text describing it.

We shall see in the edition area the user interface prototype of the WXR application (Figure 14.c) and, next to it a series of five annotations (Figure 14.d), identifiable as yellow boxes vertically aligned immediately at the right of the WRX window. These five annotations describe each mode can be relevant in the other representations in which the abbreviation of the modes is used; they are, namely, "*OFF = Switch OFF*", "*STDY = Switch for test only*", "*TST = trigger for hardware checkup*", "*WxON = activate radar detection*", and "*WXA = Focus detection on alert*". These annotations were created as *Explanations,* meaning that they are intended to explain the different modes of the WXR panel on the prototype. An additional annotation (Figure 14.e) named "*TODO: Ergonomic inspection reference*" is also present and it is attached a reference to a PDF file of the Ergonomic Criteria for the Evaluation of Human-Computer Interfaces by Christian Bastien and Scapin [43]; that reference is proposed as a recall for that user interface still requires an ergonomic inspection. The annotations created with the ARMADILLO *plugin* are automatically saved in the ARMADILLO *repository*.

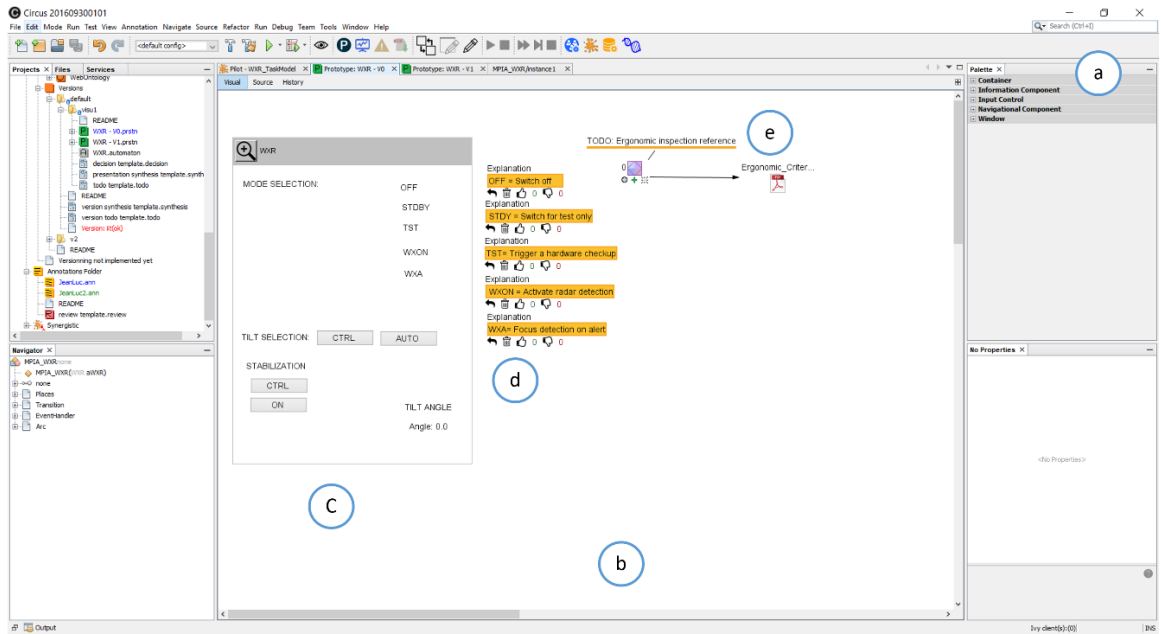

Figure 14 : Annotations on the medium-fidelity prototype for the weather radar application.

## 5.3 Importing annotations from multiple artefacts

The annotation created for a specific artefact can be reused to annotated other artefacts in the same project. Hereafter we illustrate how annotations created using the PANDA editor (describing the user interface prototype) can be associated with ICO models (describing the system behavior). We present here how these annotations can be connected together and why connecting them is relevant. To start, we should select all the annotations created with PANDA and then make them to point to the file describing the ICO system model. This is made by selecting the annotation and informing the ICO system model as a target. The next step is to launch the tool PETSHOP and open the ICO model set as target. While opening the ICO model, the ARMADILLO *plugin* looks for annotation in the *repository* that have as target the ICO model. When annotations are found, they are displayed in the edition area of PETSHOP as shown at Figure 15.

As we shall see, the annotations in PETSHOP are initially displayed at the same coordinates as they used to be in PANDA. This initial position is meant to facilitate the visual identification of annotation freshly imported. However, different representations of the interactive system might have their own layout and the annotations are unlikely to be located at the same place for each artefact. Thus, it is necessary to move the annotation to an appropriate location. Figure 16 shows the annotations that have been moved in the PETSHOP environment to establish a relationship with the corresponding areas of the ICO model. Since we cannot use the selection of fragment for the annotations, we rely on the location of the annotation to associate them with the relevant fragment of the model. The location and the size of one annotation are properties that are specific to one target while the other properties like the content are shared. Thus, moving the annotations in this model will not affect their display on the low-fidelity prototype and vice-versa. Thus, the annotations created to explain the meaning of the radio button on the user interface prototype (for example *OFF = Switch OFF"*) are now reused to explain which parts of the ICO model are responsible for describing the behavior of that button.

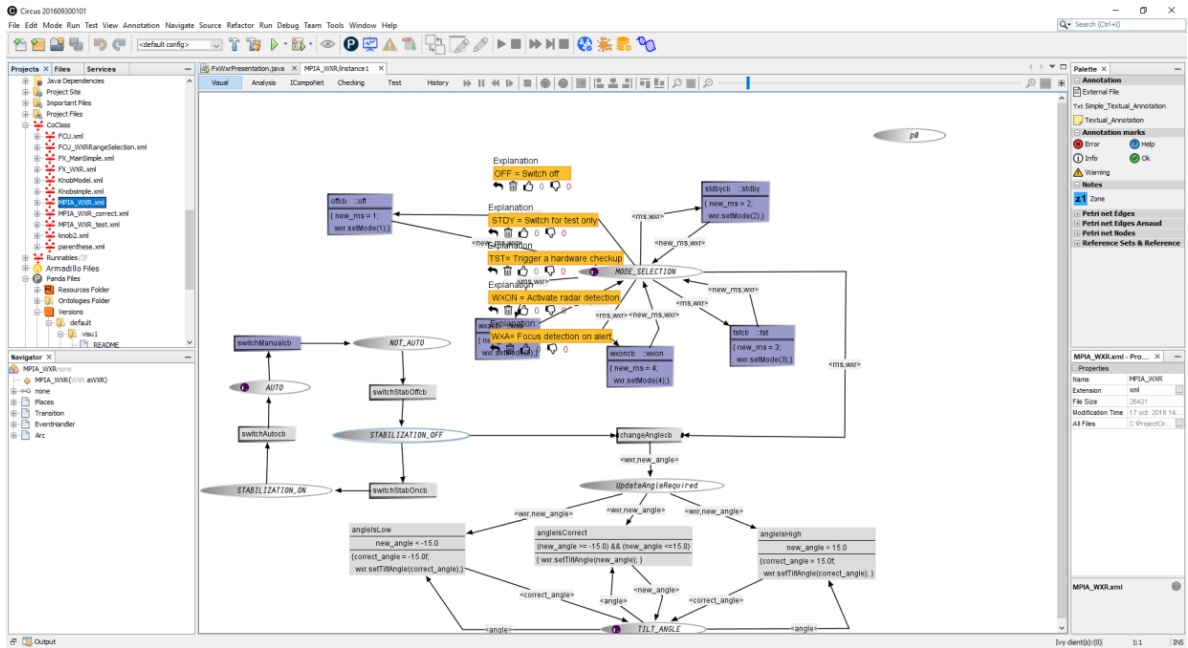

Figure 15 : Annotating an ICO model using the editor PETSHOP: initial display of annotations imported from PANDA.

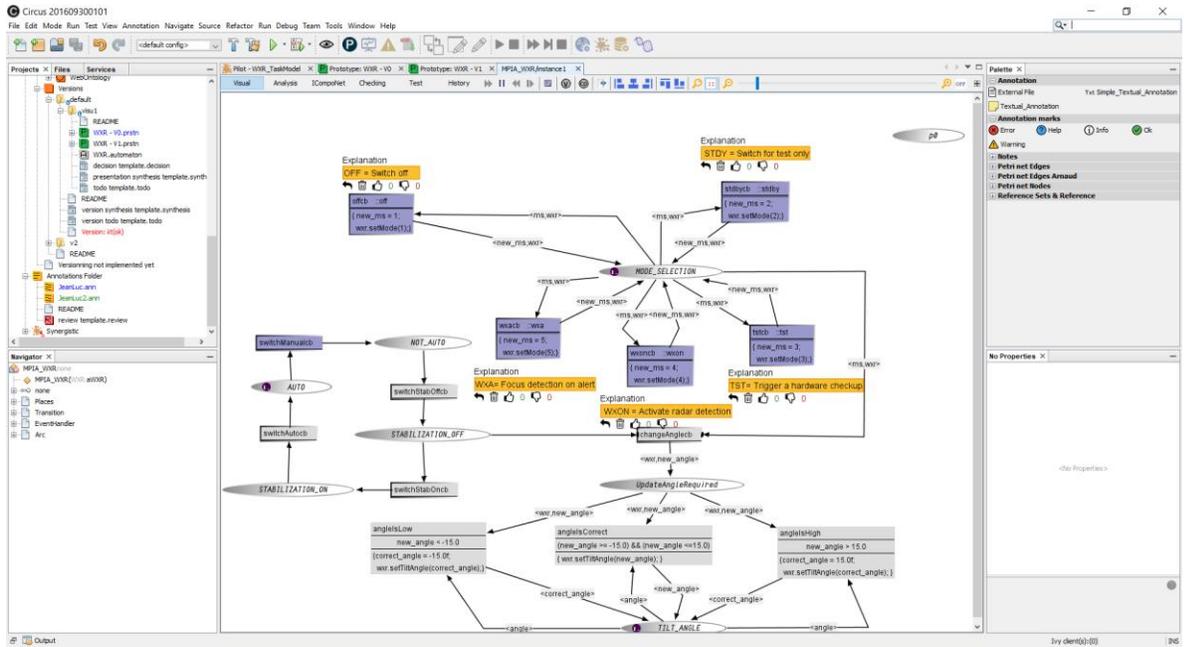

Figure 16 : Annotating an ICO model using the editor PETSHOP: repositioning annotations to target elements in the model.

## 5.4 Overview of all annotations over multiple artefacts

The ARMADILLO *plugin* presented so far provides a local view of annotations created with a particular editor and annotations that were imported from multiple artefacts. In order to have a complete view of all annotations created for the project, possibly including multiple artefacts, we have created a tool called ARMADILLO *viewer* which is shown by Figure 17. The ARMADILLO *viewer* is available at the project management level of the CIRCUS environment. It is as a standalone tool that is able to parse the files describing annotations in the repository and recreate a graphic view that retraces all annotations to targets and artefacts. The Figure 17 illustrates such as an overview of all annotations created for the case study. As we shall see, there are five annotations (namely, , "*OFF = Switch OFF*", "*STDY = Switch for test only*", "*TST = trigger for hardware checkup*", "*WxON = activate radar detection*", and "*WXA = Focus detection on alert*") originally created using PANDA and then connected to ICO models using PETSHOP. Such as connections are indicated to a pointer to the ICO file "*MPIA_WXR.xml*" and another pointer to the PANDA prototyping file "*WXR – V0.prstn*".

These annotations are graphical representations of everything available in the ARMADILLO *repository* for the project used as case study. Such as annotations contains all the metadata describing the editor, authors and classpath allowing to open the corresponding editors so each annotation in the context of artefacts being annotated.

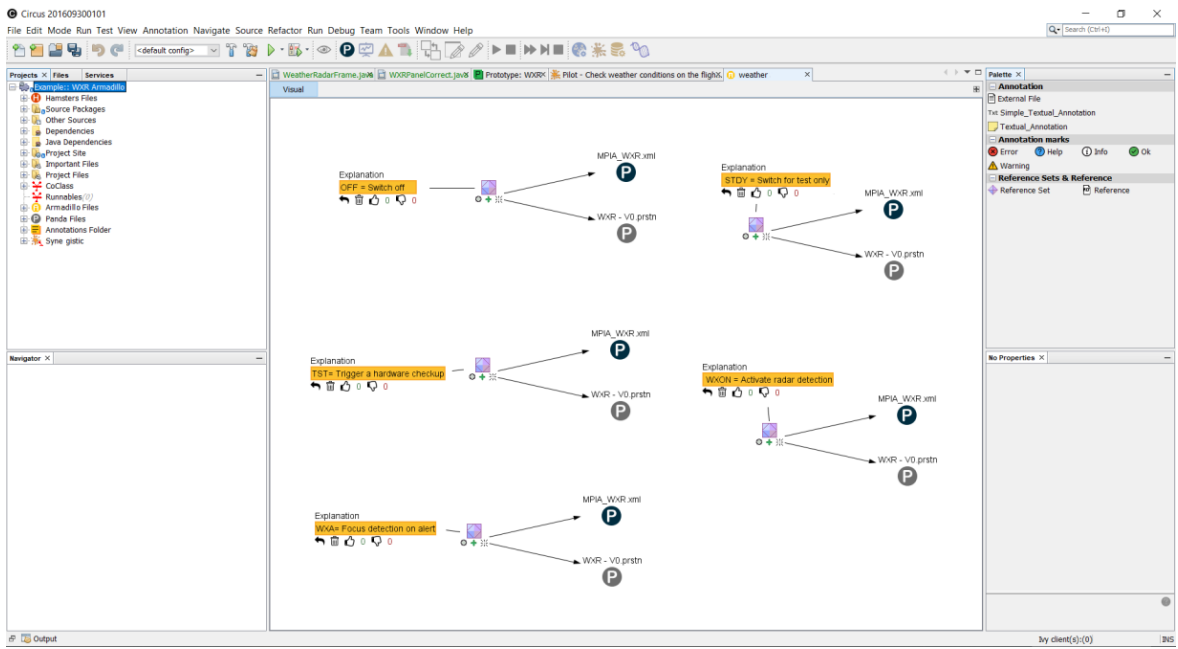

Figure 17 : Overall view of annotations over multiple artefacts using ARMADILLO.

## 5.5 Benefits of the tool from an end-user perspective

The contributions presented above were initiated by several research projects with different companies (mainly SoftTeam and Airbus). While previous sub-sections have presented the use of ARMADILLO and its underlying concepts to a small but real-life case study, this section highlights the genesis of this work and reports on feedback received from the various stakeholders and how this feedback was used to design and develop ARMADILLO.

*5.5.1 Project context*

We focus here on a set of research projects performed with different departments at Airbus. One was with interactive cockpits safety engineering group and the other one was with the Flight Warning System team. These research projects lasted over a decade and in total and resulted in publication co-authored with Airbus engineers and experts. As for the interactive cockpits formal models engineering the interested reader can refer to [52] or [53] for Flight Warning Systems modelling and [54] or [55] for formal description of interactive cockpits.

The aim of these projects was to develop a Model-Based System Engineering approach [59] dedicated to the modeling of complete set of interactive cockpits components, including graphical widgets (compatible with the ARINC 661 Specification standard [56]), interaction techniques and interactive applications (to be deployed in the new generation of interactive cockpits from Airbus manufacturer). During the recent years, the focus was on touch interactions design and modelling [57] and software architecture of interactive services connected to aircraft systems [58].

During these projects various notations and associated tools were designed and developed each of them focusing on specific behaviors (e.g. interactive system behavior [42] or operators' tasks descriptions [60]). More holistic descriptions of aircraft systems were using SysML-related modelling [58].

*5.5.2 Identification of stakeholders needs*

From the beginning of these projects it was that a quadruple problem had to be solved:
- Designing notations that have an expressive power high enough to describe all the elements that had to be described;
- Designing notations that are semantically close enough to the behaviors to be modeled so that the models would remain meaningful and would encompass modeling primitives close to the information to be modelled;
- Designing connections between the notations so that information used in one model can be referred to in another model;
- Support the editing of each kind of models with dedicated tools.

While multiple tools and notations were designed and built (presented in previous sections) e.g. PANDA for interactive applications prototypes, PetShop for interactive systems behavior and HAMSTERS for modeling operators' goals and tasks it was clear that some information would not be possible to integrate in the models. First, vernacular and informal requirements, that are produced in the early phases of the industrial project (such as early requirements for interactions to be used in aircraft cockpits) are too generic to appear in a behavioral model of a system or an operator. We use vernacular here instead of informal as engineers working for a long period of time in the same company develop a language that is exploiting a lot of acronyms and embeds the company culture. Second, requirements (due to their abstract nature) are imprecise and incomplete leaving a lot of space for interpretation. In that case, a given requirement considered in a given model might be differently integrated in another one. Third, these different views on the requirements requires discussions and argumentation in order to reach an agreement. Modeling tools usually don't integrate those aspects as argued in [61].

*5.5.3 Two concrete examples*

**Informal requirements in different artefacts: Addressing different level of DAL in different artefacts**

Figure 18 presents an envisioned application that could replace the current Flight Control Unit (FCU) panel below the windshield. That FCU panel allows multiple actions including entering parameters for the auto pilot and engaging and disengaging it. The user interface in Figure 18 is composed of two parts. The upper part of the left panel of the EFIS (Electronic Flight Information System) is dedicated to the configuration of the barometer settings. The top right panel of the EFIS page enables the display of several navigation information (such as waypoints). The value of the barometer setting is DAL A [63] according to DO 178C classification [67]. In the field of critical systems, safety standards such as DO-178C or IEC 61508 define Development Assurance Levels for software systems (or for functions of software systems). These levels are based on the analysis of consequences or effect of a malfunction. For instance, if a function failure has high consequences such as multiple fatalities, it is called catastrophic and certification authorities will require that the system manufacturer will provide a Development Assurance Level A (DO-178C standard for aeronautics [67]). If consequences are lower, the required level will decrease. Developing a system of a DAL A is extremely resource consuming and expensive and, as far as software is concerned, the use of formal description techniques is required [66]. In lower DALs, such expensive approaches are not required and for reaching levels such as DAL D rigorous software engineering approaches are sufficient

The right-hand side is less critical. As presented in [62] within the Airbus project we have proposed to engineer and model the barometer setting and to program the other part. This information was added to the ICO model that was only coping with the barometer. A question that arise was whether or not to make this information about the criticality salient or not on the user interface. In order to keep the user interface consistent with the other ones and due to the fact that the pilots are trained and know the importance of the barometer value it was decided not to change the appearance. This information was added to the user interface prototype and connected to the formal model to ensure completeness (the information is present in each artefact) and consistency (each artefact has processed the information, even though in that concrete case nothing was changed on the user interface).

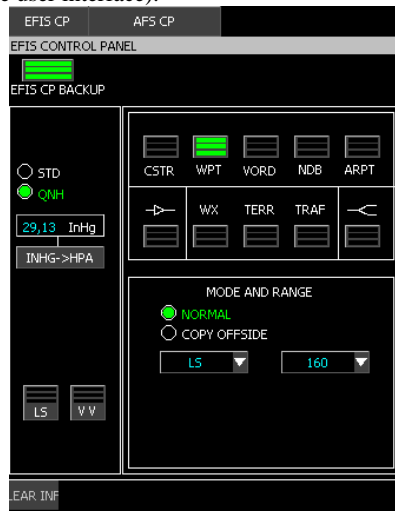

Figure 18 : EFIS control panel (left panel is highly critical DAL A while right panel is less critical DAL C [63])

**Requirements addressed in different artefacts: The mutual exclusion of buttons**

Stakeholders of the project expressed the need for describing information that was outside of the expressive power of the ICO notation [40] that was used for interactive widgets and interactive applications. The requirement was to ensure that two functions would always be triggered in a mutually exclusive way (e.g. engaging and disengaging the autopilot). A procedural state-based notation like ICOs allows easily to model such a behavior as shown in Figure 19. In the current state (Disengaged) the only transition available is Engage. Triggering that transition would set a token in place Engaged making the transition Engage unavailable and the transition Disengage available.

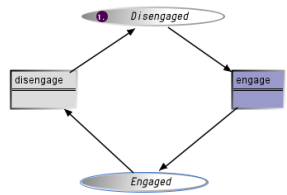

Figure 19 : Engaging and disengaging transitions which are mutually exclusive in an ICO model

While formal analysis would demonstrate that these two transitions are mutually exclusive but it is not explicit in the model that this requirement is met. Annotations were immediately added to the ICO model to record that information in an explicit manner. In addition, it was claimed important by the stakeholders that this property should be reflected and ensured on the user interface using toggle buttons (instead of two different buttons one being enabled and the other one being disabled). That property was thus appearing also at the user interface prototype artefact that was connected to the ICO formal model.

*5.5.4 Impact on the design and improvement of ARMADILLO*

As presented in the two examples above the need to offer the possibility to engineers to use annotations was made salient from the very beginning of the project. Once the need for a tool support was settled, we start to work on the design starting with paper-based prototype and ultimate leading to the design of the tools previously presented. One of the challenges issues was where to place information (edition area, list of annotation created, controls). For that, we have investigated several design opnions. The process of creation of ARMADILLO was iterative and incremental. The development of our approach took almost four years. Along the project, we had regular meetings with the users, usually every two or three months. These meeting served as formative evaluation for the design of the tools. We collected feedback through informal interviews that follow a demonstration of a prototype. Interviews were not formalized but three questions we regularly made: what do you like with the prototype? what you do not like? do you have suggestions for improvements? Unfortunately, we did not have made transcripts of answers uttered by the participants. For the need of the project at that stage, a backlog of improvements to be made was considered enough. Nonetheless, the most important requests made by the users are reported hereafter as milestones for the development of the tool. Initially, the request was to have annotations for user interface prototypes. A series of prototypes was created until we found the best placement for annotations in the palette of PANDA (as shown in Figure 9). The second milestone was to create annotations formats that suits the user needs. This was not a straightforward process. We start working with basic annotation types such as text and drawings, but through the meetings and discussions with engineers we have identified some complex annotations types (such as voting mechanisms, markers and scenarios) that trigger cycles of design and test for that particular annotations. The case of annotations for scenarios, in particular, was identified as a mean for defining non-regression. Indeed, when making changes in the prototypes, the engineers could break design solutions that worked.

So that, it was suggested to create annotations to indicate things that cannot be changed when prototypes evolve. Some studies investigating the use of annotations as testing scenarios are described in [64][65]. At this point, once we had a suitable solution for annotating user interface prototypes, we generalize the solution using a plugin that could be reused to give to other tools the functions for annotating HAMSTERS models and the ICO models using PETSHOP. The last step was the creation of the ARMADILLO repository and the viewer allowing to build the overview of all annotations in a project.

Section 5.5. of this paper has highlighted the importance for the various stakeholders involved in various research projects to provide means of annotating the various interactive systems-related artefacts produced in the design, prototyping, specification and modeling of interactive systems. Beyond the simple addition of annotations these stakeholders highlighted the importance of structuring annotations, reusing them and connecting them to one or several artefacts. The ARMADILLO tools was iteratively designed and modified to support these needs and to make complex tasks easy to perform. As the tool was developed mainly with one company, the identified needs and tasks of stakeholders might not be complete with respect to other companies practices. By making the tool publicly available we hope that more needs will be identified and in that case, we will extend ARMADILLO to cover them.

## 6 CONCLUSIONS

This paper presented a tool-supported annotation model for the design and development of interactive systems. The paper builds on related work to demonstrate the importance of annotations the development of interactive systems and proposes a generic solution (a meta-model) to engineers the use of these annotations. As demonstrated by other authors, annotations play a critical role for communicating ideas, information and decisions along the development process. The meta-model proposed addresses the multi-form nature of annotations and may be further extended to cover new needs. The ARMADILLO tool supports that meta-model and provides additional features such as versioning of annotations. In order to support annotation sharing among multiple artefacts dedicated mechanisms have been presented. This supports the entire life-cycle of an annotation but support also the design process by having annotations as a first class information and not only (as this is usual the case) a side product to be handled as a side product.

While the W3C annotation data model [16] addresses web applications our extended version of it is specially targeting at interactive systems and supports the User Centered Design process of ISO 9241 part 210 iterative process.

Beyond the model and the tool ARMADILLO that supports it; we have presented their integrability and integration in the existing tools in the CIRCUS platform [40] (which encompass the tools PANDA, HAMSTERS, and PETSHOP). This is done by a generic plugin that has been instantiated to every tool editor of CIRCUS platform and used to produce specific artefacts (e.g. task models or dialog models). While the code of our tools are not open source (industrial constraint), the ARMADILLO and the other editors in the CIRCUS platform are available and can be used by the community free of charge.

So far the plugin has been deployed in a few editors of CIRCUS but the meta-model and the plugin concept are generic enough to be deployable in other tools. In order to support design and development activities, the plugin support the entire life cycle of annotations including their connection to the different artefacts produced by different stakeholders involved in the design process of interactive systems. On one hand, the choice of a plugin architecture is justified by some of the inner advantages such as consistent interaction with the annotation tool across multiple editors, reuse of the code, extensibility of annotations to other editors, etc. On the other hand, plugin are not a panacea and are tied to a specific development platform. For that we cannot claim that the plugin tools presented in this paper are an universal solution. However, our approach illustrated by the means of the tools (including centralized repository for annotations, using standards for describing annotations with extensions for metadata and annotation types, dedicated tools for tracking the annotations distributes along multiple artefacts, etc.) can be used as an example to build other similar tools in other environments to solve some problems such as : how to systematically enrich and bind annotations to digital artefacts/models used to specify an interactive systems,  thus supporting decision making along the development process; how to cross-reference and cross-check design decisions to multiple artefacts; how to follow the evolution and ensure the consistency of design decisions along the development of interactive systems.

Relationship with annotations in programming languages (R1). We can easily expand the section state-of-the-art to encompasses the uses of annotations in this context as we did the work as part of the industrial project. However, we decided not to include this in the current paper as the case study is only dealing with graphical notations. We propose to add a short paragraph to the state of the art explaining that our annotation model covers current practice in programming.

We have demonstrated the use of the annotation models and the ARMADILLO tool in a case study on the aerospace domain involving four different artefacts. The case study demonstrates that it is possible to collect several information that might be useful for supporting design decisions. This opens-up several perspectives for investigating the use of informal information collected and design decisions along the process. In future work, we will focus on traceability of the decisions and tools for supporting the visualization and rational design. Now that the concept is operational, we have the tools for a large case study including the collection of data describing the results of usability testing.

**ANNEX I – List of annotations tool examined for the state of the art on annotations**

| Year | Name of the tool | URL | Type |
|------|------------------|-----|------|
| 1988 | Quilt | https://dl.acm.org/citation.cfm?id=62282 <br> https://dl.acm.org/citation.cfm?id=45414 | A |
| 1988 | Amaya | https://www.w3.org/Amaya/ <br> https://dev.w3.org/Amaya/doc/WX/Annotations.html | A |
|      | Annozilla / Annotea project | http://annozilla.mozdev.org/ | C |
| 2005 | Diigo | https://www.diigo.com/ | C |
| 2007 | sense.us | http://vis.stanford.edu/papers/senseus | A |
| 2008 | Protonotes | http://www.protonotes.com/ | C |
| 2009 | SparTag.us | https://dl.acm.org/citation.cfm?id=1385582 | A |
| 2009 | HyperImage 3/ | http://hyperimage.ws/en/ | A |
| 2010 | D.note | https://dl.acm.org/citation.cfm?doid=1753326.1753400 | A |
| 2011 | List-it | https://dspace.mit.edu/handle/1721.1/73002 | A |
| 2011 | LiquidText | https://dl.acm.org/citation.cfm?id=1979430 | A |
| 2011 | Zydeco | https://dl.acm.org/citation.cfm?id=1979038 | A |
| 2011 | Annotatorjs | http://annotatorjs.org/ | C |
| 2012 | Elias' « unnamed » | https://dl.acm.org/citation.cfm?id=2208288 | A |
| 2012 | ChronoViz | https://dl.acm.org/citation.cfm?id=2208590 | A |
| 2012 | GatherReader | https://dl.acm.org/citation.cfm?id=2208327 | A |
| 2012 | Annotorious | https://annotorious.github.io/ | A |
| 2012 | Domeo | https://jbiomedsem.biomedcentral.com/articles/10.1186/2041-1480-3-S1-S1 | A |
| 2014 | Instant Annotation | https://dl.acm.org/citation.cfm?id=2557209 | A |
| 2014 | Neonion | https://dl.acm.org/citation.cfm?id=2699012 | A |
| 2017 | Dokieli | http://csarven.ca/dokieli-rww | A |
| 2017 | Hypothesis | https://web.hypothes.is/ | C |
| 2017 | Authorea | https://www.authorea.com/ | C |
| 2017 | Pundit Annotator | http://net7.github.io/pundit2/ | A |
| 2017 | Ponga | https://www.ponga.com/ | C |

**Legende**: **A** academic tool, **C** commercial tool.